\documentclass[journal=langd5,manuscript=article]{achemso}

\usepackage{graphicx}
\usepackage{amsmath,amssymb,amsfonts}

\newcommand{\eg}{{e.g., }}
\newcommand{\ie}{{i.e., }}

\newcommand{\vecf}{{\bf f}}

\newcommand{\vecq}{{\bf q}}
\newcommand{\vecr}{{\bf r}}

\newcommand{\vecu}{{\bf u}}
\newcommand{\vecv}{{\bf v}}

\newcommand{\boldRho}{\boldsymbol{\rho}}
\newcommand{\lec}{l_{\rm ec}}
\newcommand{\etab}{\eta_{\rm b}}
\newcommand{\Gb}{G_{\rm b}}
\newcommand{\GL}{{\cal G}_{\rm L}}
\newcommand{\GP}{{\cal G}_{\rm P}}
\newcommand{\GT}{{\cal G}_{\rm T}}

\author{Chen Bar-Haim}
\email{chentush@gmail.com}

\author{Haim Diamant} 
\email{hdiamant@tau.ac.il} 
\affiliation{Raymond \& Beverly Sackler School of Chemistry, Tel Aviv
University, Tel Aviv 6997801, Israel}

\title{Surface response of a polymer network:\\
  Semi-infinite network}

\begin{document}

\setlength{\baselineskip}{8pt}
\setlength{\parskip}{6pt}

\begin{abstract}
\setlength{\baselineskip}{6pt}
  
We study theoretically the surface response of a semi-infinite
viscoelastic polymer network using the two-fluid model. We focus on
the overdamped limit and on the effect of the network's intrinsic
length scales. We calculate the decay rate of slow surface
fluctuations, and the surface displacement in response to a localized
force. Deviations from the large-scale continuum response are found at
length scales much larger than the network's mesh size. We discuss
implications for surface scattering and microrheology. We provide
closed-form expressions that can be used for surface
microrheology\,---\,the extraction of viscoelastic moduli and
intrinsic length scales from the motions of tracer particles lying on
the surface without doping the bulk material.
\end{abstract}


\vspace{-0.5cm}
\section{Introduction}
\label{sec_intro}

Surfaces of materials exhibit distinctive behaviors compared to the
bulk \cite{SurfacesAdamson}. This applies, in particular, to the
deformations of liquid and solid surfaces in response to stresses and
thermal fluctuations, as has been studied for many years. Important
examples are capillarity of liquid surfaces \cite{CapillarityWetting}
and Rayleigh waves on solid surfaces \cite{LLelasticity}.  In the
present work we consider the surfaces of materials whose response lies
in-between these two limits, i.e., viscoelastic media made of both
fluid and solid components.

Surface waves in polymer solutions and gels have been thoroughly
studied using various light scattering and mechanical excitation
techniques
\cite{Cao1991,DorshowPRL1993,MonroyLangevin1998,OnoderaChoi1998,Yoshitake2008}. (See a recent review in ref~\citenum{MonroyReview2017}.) A characteristic
feature of these experiments is the crossover from capillary
(surface-tension-dominated) waves to Rayleigh (elasticity-dominated)
waves with decreasing frequency or increasing concentration. The
prevalent theory for the dispersion relation of these waves on the
surface of a semi-infinite viscoelastic medium is by Harden, Pleiner,
and Pincus \cite{HardenPincus1st,Pincus,Pleiner1988}. Its
validity has been confirmed by the experiments mentioned
above. Extensions have been developed for more complicated scenarios,
such as the existence of an adsorbed layer of different mechanical
properties \cite{KapplerNetz2017,HuangWang1997}, and a supported film
of finite thickness \cite{HenleLevinePRE2007}. All these models have
considered the strong coupling limit where the viscous and elastic
components move together as a structureless medium.

Our focus here is different. We are interested in the effect of the
intrinsic length scales characterizing a structured fluid on its
surface response, and how these characteristics can be extracted from
surface measurements. This is motivated by recent studies, which have
revealed a dynamic length scale intermediate between the material's
static correlation length and the limit of structureless bulk response
\cite{ViscoelasticIntermediate,DynamicCorrelationLength,HaimEPJE,Granek2018}.
The intermediate length scale was found to be associated with a
distinctive mechanical response. Therefore, we extend earlier theories for
the relaxation of liquid and solid surfaces \cite{Pershan2012}, as
well as the response of the surface to a localized force
\cite{LLelasticity}, to the case of a structured viscoelastic medium
described by the two-fluid model \cite{Levine2Fluid2001,
  HaimEPJE,DeGennes1976,DeGennes1976II,DoiOnuki1992,Milner1993,ChromatinRabin}.

The present work is particularly relevant to the study of the
mechanical response of soft materials. Viscoelastic response is
commonly studied by rheometers \cite{LarsonBook}, but this macroscopic
technique cannot be applied directly to the surface of the
material. (Quasi-two-dimensional rheometers were developed for thin
layers lying at fluid interfaces \cite{Dennin}.) One alternative is to
extract the viscoelastic moduli from the dispersion relation of
surface waves \cite{MonroyReview2017}.
Another relevant technique is microrheology, where the viscoelastic
moduli are extracted from the motions of tracer particles
\cite{MasonWeitz,TwoPointCrocker,MicrorheologyReview}. This technique
has been applied primarily in the bulk. It may be advantageous in
certain cases to obtain bulk microrheological properties from the
motions of tracer particles positioned on the surface of a material,
without doping the bulk
\cite{LadamSackmann,expDennin2012,expDennin2013,expDennin2014,KomuraMaterials2012,KomuraEPL2012}. As
far as we know, there is no theory to accompany such experiments, which
considers the material as a structured fluid.

The article is organized as follows. After presenting the model, we
divide the results into two parts. The first concerns the overdamped
dispersion relation (decay rate) of fluctuation modes on the surface
of a viscoelastic medium. In the second part we study the surface
deformation of such a medium in response to a localized
force. Finally, we discuss the findings and their potential usage in
experiments.


\vspace{-0.5cm}
\section{Model}
\label{sec_model}

\begin{figure}
  \centerline{\includegraphics[width=0.6\textwidth]{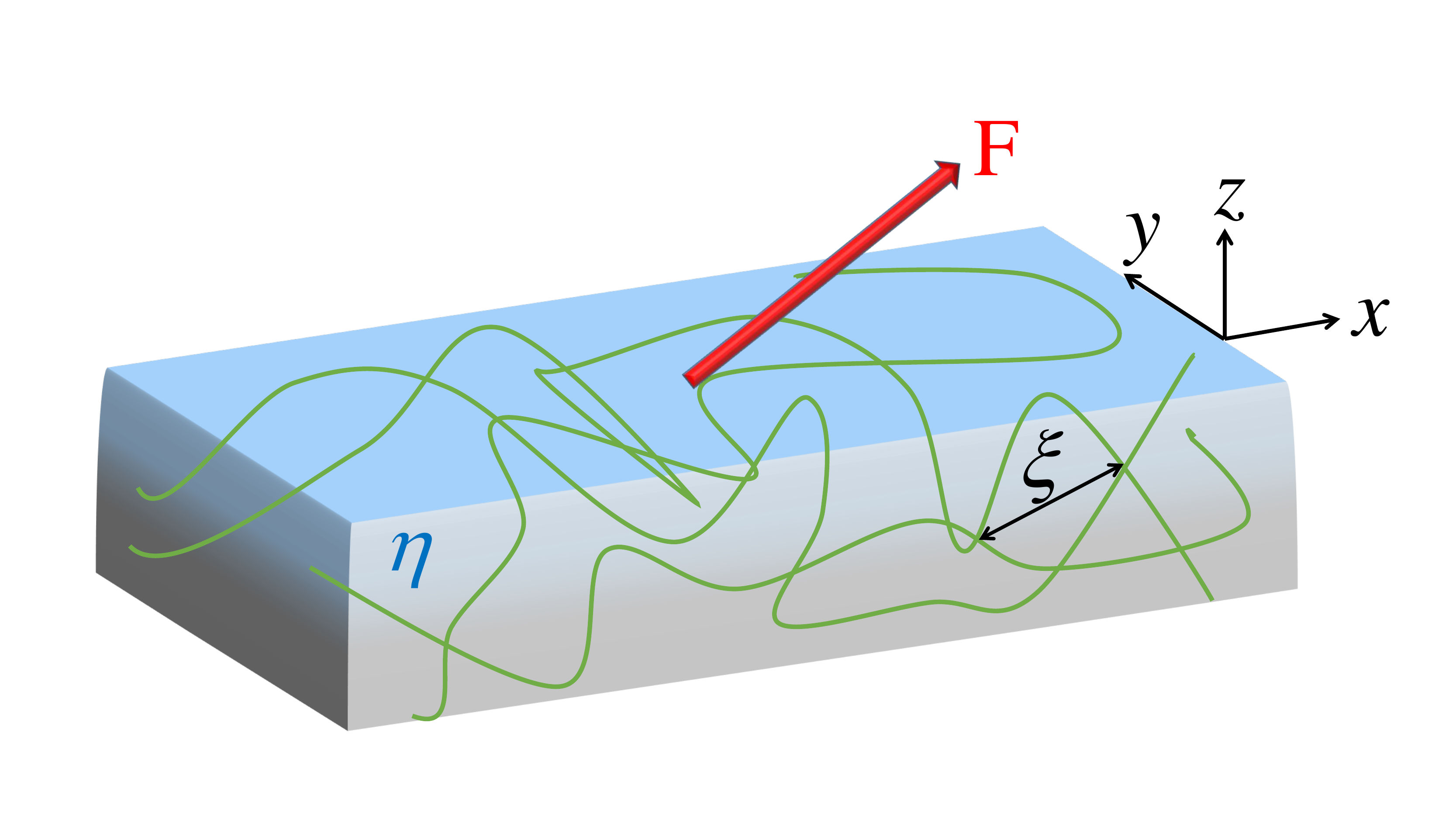}}
\caption{\setlength{\baselineskip}{6pt}
    Schematic view of the system and its parameters.}
\label{fig_scheme}
\end{figure}

We consider a semi-infinite polymer solution, occupying the region
$z<0$. We use the two-fluid model \cite{Levine2Fluid2001,HaimEPJE,DeGennes1976,DeGennes1976II,DoiOnuki1992,Milner1993,ChromatinRabin}
to describe the structured medium.
The model has two components --- a semi-dilute polymer network,
structurally characterized by a correlation length $\xi$, and a
structureless solvent. See the schematic illustration in
Figure~\ref{fig_scheme}. The network is described as a (visco)elastic
medium, whose deformation is defined by a displacement field
$\vecu\left(\vecr,\omega\right)$, which is a function of position
$\vecr=(\boldsymbol{\rho},z)$ and frequency $\omega$. The
corresponding stress tensor is
\begin{equation}
\sigma_{ij}^{(u)}=2G\left[u_{ij}-\left(u_{kk}/3\right)\delta_{ij}\right]+Ku_{kk}\delta_{ij},
\label{StressTensorElasticNet}
\end{equation}
where $u_{ij}\equiv\left(\partial_iu_j+\partial_ju_i\right)/2$ is the
network's strain tensor, and $G$ and $K$ its shear and compression
moduli, which may be frequency-dependent. We distinguish between the
shear modulus of the {\em bare} network, $G(\omega)$, and the modulus
of the bulk material, $\Gb(\omega)$, to be presented below. The
solvent is described as a viscous incompressible fluid, having a flow
velocity field $\vecv\left(\vecr,\omega\right)$, pressure field
$p\left(\vecr,\omega\right)$, and the stress tensor
\begin{equation}
\sigma_{ij}^{(v)}=-p\delta_{ij}+2\eta v_{ij},
\end{equation}
where $v_{ij}\equiv\left(\partial_iv_j+\partial_jv_i\right)/2$ is the
fluid's strain-rate tensor, and $\eta$ its shear viscosity.

The two components are coupled through mutual friction characterized
by a coefficient $\Gamma$.  In the present work, we focus on the
overdamped response of a surface. Thus, neglecting inertia, we write
the governing equations for the three fields
$(\textbf{u},\textbf{v},p)$ as follows:
\begin{eqnarray}
&&0=\nabla\cdot\sigma_{ij}^{(u)}-\Gamma\left(i\omega\vecu-\vecv\right),\label{eq_motion_net}\\
&&0=\nabla\cdot\sigma_{ij}^{(v)}-\Gamma\left(\vecv-i\omega\vecu\right),\label{eq_motion_fluid}\\
&&0=\nabla\cdot\vecv. \label{eq_incomp}
\end{eqnarray}
The first two equations, together, reflect the conservation of
momentum in the composite material. The third accounts for the
conservation of mass which, in the assumed limit of a semi-dilute
solution, is dominated by the incompressible solvent.  The frictional
force density in eqs~\ref{eq_motion_net} and \ref{eq_motion_fluid} is
proportional to the local relative velocity of the two components,
thus maintaining overall Galilean invariance.  The friction
coefficient $\Gamma$ is related to the correlation length $\xi$ as
$\Gamma\sim\eta/\xi^2$ \cite{HardenPincus1st}.

Equations \ref{StressTensorElasticNet}--\ref{eq_incomp} are
supplemented by the following boundary conditions.  (a) Far deep into
the medium both components are stationary,
\begin{equation}
  \vecu\left(\boldRho,z\rightarrow-\infty\right)=
  \vecv\left(\boldRho,z\rightarrow-\infty\right)=
  \nabla p\left(\boldRho,z\rightarrow-\infty\right)=0.
\label{bc_infty}
\end{equation} 
(b) At the surface the two components move together,
\begin{equation}
	i\omega\vecu(\boldRho,z=0)=\vecv(\boldRho,z=0).\label{bc_freeSurface1}
\end{equation}
This boundary condition implies strong coupling between the two
components at the surface. We discuss it further in the
Discussion section, where we also offer an alternative if this
assumption should be relaxed.
(c) At the surface the total stress
in the medium balances the other surface forces,
\begin{equation}
  \sigma_{iz}^{(u)}\left(\boldRho,z=0\right)+\sigma_{iz}^{(v)}\left(\boldRho,z=0\right)=
  f_i-\gamma\nabla^2_{\rho}u_z\delta_{iz}.\label{bc_freeSurface2}
\end{equation}
Here, we have included an external surface force density $\textbf{f}$
and a restoring force due to the surface tension $\gamma$, where
$\nabla^2_{\rho}$ is a two-dimensional laplacian.  Equations
\ref{StressTensorElasticNet}--\ref{bc_freeSurface2} define our
model.

It is convenient to introduce the following parameters to make the
physics more transparent:
\begin{eqnarray}
&&\etab\equiv G/\left(i\omega\right)+\eta,\\
&&\xi\equiv\left(\frac{G\eta}{i\omega\Gamma\etab}\right)^{1/2},\\
  &&\lambda\equiv\left(\frac{K+4G/3}{i\omega\Gamma}\right)^{1/2}=
  \left[\frac{2\left(1-\nu\right)}{1-2\nu}\frac{\etab}{\eta}\right]^{1/2}\xi,
\label{charac_lengths_lambda}
\end{eqnarray}
where $\nu\equiv (3K-2G)/[2(3K+G)]$ is the network's Poisson ratio.
The bulk viscosity $\etab(\omega)$ characterizes the large-scale shear
response of the two-component medium. We note that it emerges from the
smaller-scale parameters, $G$ and $\eta$, and is complex in principle,
containing both elastic and viscous contributions. The material's bulk
shear modulus is simply $\Gb=i\omega\etab$. The correlation length
$\xi$ is another emergent property. It is known to coincide (up to a
factor of order unity) with the network's mesh size
\cite{ViscoelasticIntermediate,DynamicCorrelationLength}. It also
characterizes the spatial decay of transverse (shear) stresses due to
the friction between the two components and, therefore, decreases with
increasing $\Gamma$. An important conclusion is that any experiments
tapping into the internal structure of the complex fluid require a
finite network-solvent friction $\Gamma$ to be accounted for by the
two-fluid model. For the low frequencies assumed here, we expect a
small contribution to $\Gb$ from the solvent, \ie $\Gb(\omega) \simeq
G(\omega)$; then, $\xi^2 \simeq \eta/\Gamma$ is insensitive to
frequency, as physically expected and as noted above. Another
length, $\lambda>\xi$, is related to the compressive (longitudinal)
response of the polymer component. Typically $\etab\gg\eta$ and,
hence, $\lambda\gg\xi$. Moreover, $\lambda$ diverges in the limit of
an incompressible network ($\nu\rightarrow 1/2$ or $K\rightarrow
\infty$). Note that the continuum theory that we employ is valid only
over distances much larger than $\xi$.

The surface tension together with the correlation length of the
structured fluid introduces a characteristic relaxation time,
\begin{equation}
\Omega^{-1}\equiv\frac{\eta\xi}{\gamma}
\label{char_freq}.
\end{equation}
The surface tension is associated also with an elasto-capillary length,
\begin{equation}
\lec\equiv\frac{\gamma}{i\omega\etab}=\frac{\gamma}{\Gb}
\label{eq_def_lec}
\end{equation}

\vspace{-0.5cm}
\section{Results}

\vspace{-0.5cm}
\subsection{Structure of the general solution}

Applying a Fourier transform, $\tilde{g}(\vecq,\omega)\equiv\int
g(\boldRho,\omega)e^{-i\vecq\cdot\boldRho}d^2\rho$, to all the fields
in eqs \ref{StressTensorElasticNet}--\ref{bc_freeSurface2} turns those
partial differential equations into ordinary differential ones,
dependent on $z$.  The general solution \cite{Note2}
is a composition of six modes, $e^{\pm\kappa_n z}$, $n=1,2,3$, where
\begin{equation}
\kappa_1=q,\\\
\kappa_2=\sqrt{q^2+\xi^{-2}},\\\
\kappa_3=\sqrt{q^2+\lambda^{-2}}.\label{eq_modes}
\end{equation}
The general solution for the network displacement
$\vecu(\vecq,\omega)$ contains all six modes, that for the solvent
velocity $\vecv(\vecq,\omega)$ four modes ($n=1,2$), and the one for
the solvent pressure $p(\vecq,\omega)$ four modes ($n=1,3$). Thus,
altogether, the general solution contains 14 amplitudes. These are
determined by the thirteen boundary conditions defined in eqs
\ref{bc_infty}--\ref{bc_freeSurface2}, plus the incompressibility
constraint of eq \ref{eq_incomp}. After omitting the three modes which
diverge at $z\rightarrow -\infty$, we are left with three decaying
modes, whose ``penetration depths" are defined in eq \ref{eq_modes},
and seven amplitudes to be determined. For a large-wavelength surface
perturbation ($q\rightarrow 0$), one of the modes ($n=1$) extends
throughout the medium, whereas the other two remain localized at the
surface, with penetration depths $\xi$ and $\lambda$. We note that the
two surface-localized modes ($n=2,3$) differ from the ones
discussed in ref~\citenum{OhmasaYao2011} in that they depend on the
intrinsic lengths $\xi$ and $\lambda$.

\vspace{-0.5cm}
\subsection{Dispersion relation of overdamped surface fluctuations}
\label{sec_DR}

To obtain the decay rate of the overdamped surface fluctuations as a
function of wavevector $q$, we set $\vecf=0$ in the surface
stress, eq \ref{bc_freeSurface2}, and look for nontrivial solutions
for $\vecu$, $\vecv$ and $p$. This amounts to the requirement that the
matrix corresponding to the linear equations for the seven amplitudes
should have a zero determinant. The resulting equation gives the
dispersion relation, $-i\omega(q)$.

The full equation is complicated and can be found in the supplemental
material \cite{supplemental}. It has two linear asymptotes, at small and large
$q$, with a crossover at $q\sim\lambda^{-1}$,
\begin{equation}
-i\omega=\left\{\begin{array}{ll} \frac{\gamma
  q}{2\etab},\ \ \ \ \ \ &q\ll\lambda^{-1},\\
  \frac{\gamma q}{2\etab}\frac{\beta^2}{\beta^2-1},&\lambda^{-1}\ll q\ll\xi^{-1},
	\end{array}\right.
\label{eq_DR_result}
\end{equation}
where
\begin{equation}
\beta(\nu)=\left[2(1-\nu)/(1-2\nu)\right]^{1/2} \geq 2/\sqrt{3}.
\label{eq_def_beta}
\end{equation}
(The lower bound for $\beta$ comes from the validity range of the
Poisson ratio, $-1\leq\nu\leq 1/2$.)  The small-$q$ limit corresponds
to the bulk response of the medium and coincides with the dispersion
relation for overdamped surface fluctuations of a viscous fluid whose
viscosity is replaced by $\etab$. The other asymptote describes the
dispersion relation in the intermediate region $\lambda^{-1}\ll
q\ll\xi^{-1}$.

Since the bulk viscosity $\etab(\omega)$ is a function of $\omega$,
the equation that we obtain for the dispersion relation, along with
its asymptotes eq \ref{eq_DR_result}, in fact, is an implicit equation
for the rate, which can be solved only if $\etab(\omega)$ is known. We
give three simple limits where the dispersion relation can be obtained
explicitly.

\begin{figure}[h]
\centerline{\includegraphics[width=0.5\textwidth]{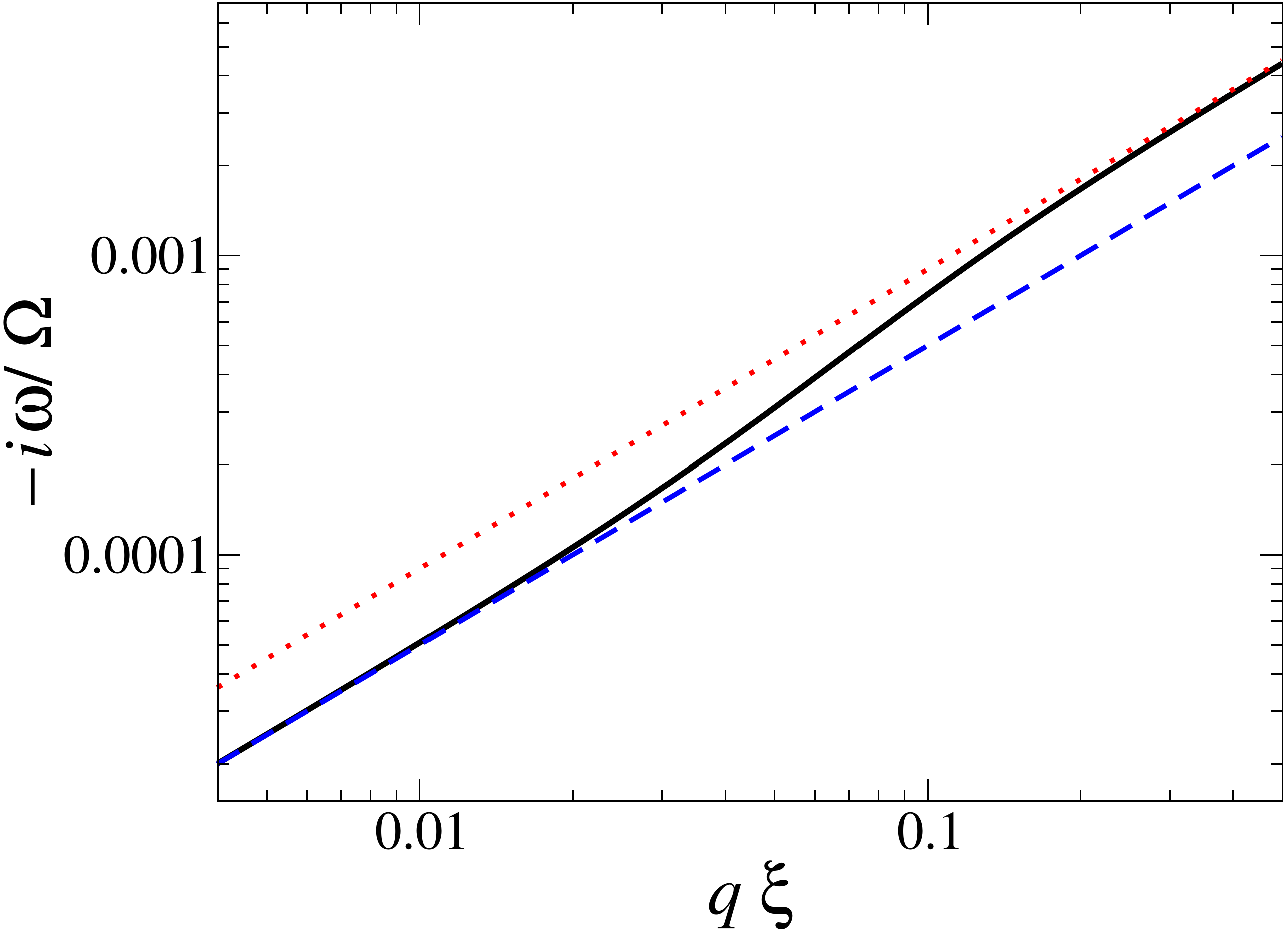}}
\caption{
\setlength{\baselineskip}{6pt}
  Normalized rate $-i\omega(q)$ as a function of normalized
  wavevector. Once normalized, the curve depends only on the network's
  Poisson ratio $\nu$ and $\etab/\eta$. We have used $\nu=0.1$ and
  $\etab/\eta = 100$. The two asymptotes given in eq
  \ref{eq_DR_result} are shown as dashed blue and dotted red lines,
  with a crossover around $q\sim\lambda^{-1}\ll\xi^{-1}$. The small
  Poisson ratio, corresponding to an unrealistically compressible
  network, is chosen to graphically emphasize the difference between
  the two regimes on a logarithmic scale. For a realistic value of
  $\nu=0.4$ we get a difference of about
  $20\%$. Equation~\ref{eq_DR_result} can be used to calculate the
  effect for any value of $\nu$. The value of $\etab$ is motivated by
  the actin networks of ref~\citenum{ViscoelasticIntermediate}, whose
  low-frequency response is governed by the viscous term, with
  viscosity about 100 times that of water.}
\label{fig_DR}
\end{figure}

The first simple case holds if $\etab$ changes slowly with $\omega$
over the range of $q$ of interest, and can be assumed constant. The
full solution in this limit is represented by the solid curve in
Figure~\ref{fig_DR}. The two asymptotes are presented as well. The gap
between the two asymptotes depends on the compressibility of the bare
polymer network, and vanishes in the limit of incompressibility. For
the curve in Figure~\ref{fig_DR} we have chosen an unrealistically
small value for the Poisson ratio ($\nu=0.1$) to demonstrate the gap
between the two asymptotes (amounting in this case to a factor of about
3). For a much less compressible network, the gap is smaller but still
not negligible; for example, for a realistic value of $\nu=0.4$ the
difference between the asymptotes is by a factor of $1.2$.

\vspace{0.5cm}

The second simple limit is where the bare polymer network is taken as
purely elastic. In this case $\etab=G/(i\omega)+\eta$, and $G$ and
$K$ are frequency-independent constants. Substitution in eq
\ref{eq_DR_result} and solving for the rate yields
\begin{equation}
  G,K=\mbox{const}:\ \ -i\omega=\left\{\begin{array}{ll} \Omega_\gamma(q)
  + \Omega_{\rm el},\ \ \ \ \ \ &q\ll\lambda^{-1},\\
  \mbox{the same, with } \Omega_\gamma \rightarrow \Omega_\gamma \beta^2/(\beta^2-1),
  &\lambda^{-1}\ll q\ll\xi^{-1}.
	\end{array}\right.
\end{equation}
We have defined here the rates $\Omega_\gamma(q)\equiv\gamma q/(2\eta)$
and $\Omega_{\rm el}\equiv G/\eta$.

A third limit which can be simply treated is where the bare network's
shear modulus has a single-exponential relaxation with some relaxation
time $\tau$, $G(\omega)=G_0/(1+i\omega\tau)$. Once the frequency
dependence of $\nu$ is neglected, it is straightforward to obtain from
eq~\ref{eq_DR_result} the explicit dispersion-relation asymptotes,
\begin{eqnarray}
  && G=G_0/(1+i\omega\tau),\,K=\mbox{const}: \\
  &&-i\omega_\pm = \left\{\begin{array}{ll}
  \frac{1}{2} \left[ \tau^{-1} + \Omega_\gamma \pm \sqrt{
      (\tau^{-1} + \Omega_\gamma)^2 - 4\tau^{-1}(\Omega_\gamma + \Omega_{\rm el})}
    \right],\ \ \ \ \ \ &q\ll\lambda^{-1},\\
  \mbox{the same, with } \Omega_\gamma \rightarrow \Omega_\gamma \beta^2/(\beta^2-1),
  &\lambda^{-1}\ll q\ll\xi^{-1}.
	\end{array}\right. \nonumber
\end{eqnarray}
This relation shows two branches, which may contain an imaginary
(oscillatory) component for sufficiently large $\Omega_{\rm el}$.

The dispersion relation of polymer networks was studied previously by
Harden et al. \cite{HardenPincus1st,Pincus,Pleiner1988}. That work
focused on a different regime, including inertia and restricting the
discussion to the strong coupling limit ($\Gamma\rightarrow\infty$ or
$\xi\rightarrow 0$).  In this limit, the material is a structureless
viscoelastic medium. We focus here on structural effects (finite
$\xi$) within the overdamped regime. The two dispersion relations
coincide only in one limit, when inertia is taken to zero in
refs~\citenum{HardenPincus1st, Pincus,Pleiner1988}, and the bulk limit
is taken in our theory ($q\ll\lambda^{-1}$). The experimental
conditions under which the present theory is valid will be examined in
the Discussion section.

\vspace{-0.5cm}
\subsection{Response to a localized surface force}
\label{sec_response}

We consider a point force, $\textbf{F}\delta(\boldRho)\delta(z)$,
applied on the surface of the semi-infinite polymer network at the
origin (see Figure \ref{fig_scheme}). Our purpose is to find the
resulting surface displacement and flow.  This can be viewed as an
extension of the elastic Boussinesq problem (ref~\citenum{LLelasticity},
Section 8) to a viscoelastic material.  The equations and boundary
conditions are given in the Model section. They are the same as those
used to find the dispersion relation in the preceding sub-section,
except that now we substitute in the boundary condition of
eq~\ref{bc_freeSurface2} a non-zero surface force density,
$\textbf{f}=\textbf{F}\delta(\boldRho)$. Applying the boundary
conditions results in this case in an inhomogeneous set of
differential equations, whose solution gives the surface values of the
different fields, $\vecu(\boldRho,z=0,\omega)$,
$\vecv(\boldRho,z=0,\omega)$, and $p(\boldRho,z=0,\omega)$. We focus
on the displacement and flow responses, which are captured by a tensor
${\cal G}$,
\begin{equation}
v_i(\boldRho,z=0,\omega)=i\omega u_i(\boldRho,z=0,\omega)={\cal G}_{ij}(\boldRho,\omega)F_j,
\end{equation}
where summation over repeated indices is used. The first equality
follows from the boundary condition of eq~\ref{bc_freeSurface1}.

The full expressions in real space could be calculated only
numerically. See more details in the supplemental material
\cite{supplemental}. The results are presented in the figures below as
solid curves. As in the preceding sub-section, we provide closed-form
expressions for the relevant asymptotic limits
\cite{supplemental}. From now on we omit for brevity the mention of
$z=0$ and $\omega$ in the arguments of the various functions. In the
expressions below, recall that $\etab$, $\beta$, and $\lec$ are in
principle all functions of $\omega$.

For the tangential response, ${\cal G}_{ij}$ with $i,j=x,y$, we find at large
distances,
\begin{equation}
\label{Gijr_eq_far}
	\rho\gg\lambda:
        \ \ \ \ \ {\cal G}_{ij}(\boldRho)=\frac{1}{4\pi\etab
          \rho}\left(\delta_{ij}+\frac{\rho_i\rho_j}{\rho^2}\right).
\end{equation}
This result coincides with the one obtained in the Boussinesq problem
\cite{LLelasticity}, $u_i=({\cal G}_{ij}/i\omega)F_j$, once
$i\omega\etab$ is replaced by the elastic solid's shear modulus, and
the solid is taken as incompressible ($\nu=1/2$). At shorter distances
we find
\begin{eqnarray}
\label{Gijr_eq_near}
	\rho\ll\lambda:
        \ \ \ \ \ {\cal G}_{ij}(\boldRho) &=& \frac{1}{4\pi\etab
          \rho}\left(A\delta_{ij}+B\frac{\rho_i\rho_j}{\rho^2}\right),
        \\
        A &=& \frac{\beta^2\etab + \eta}{(\beta^2-1)\etab+2\eta}, \nonumber\\
        B &=& \frac{(\beta^2-2)\etab + 3\eta}{(\beta^2-1)\etab+2\eta}. \nonumber
\end{eqnarray}
Note that expressions \ref{Gijr_eq_far} and \ref{Gijr_eq_near} are
independent of the surface tension $\gamma$. This is because the
surface tension acts as a restoring force in the perpendicular
direction only.

\begin{figure}[h]
\centerline{\includegraphics[width=0.45\linewidth]{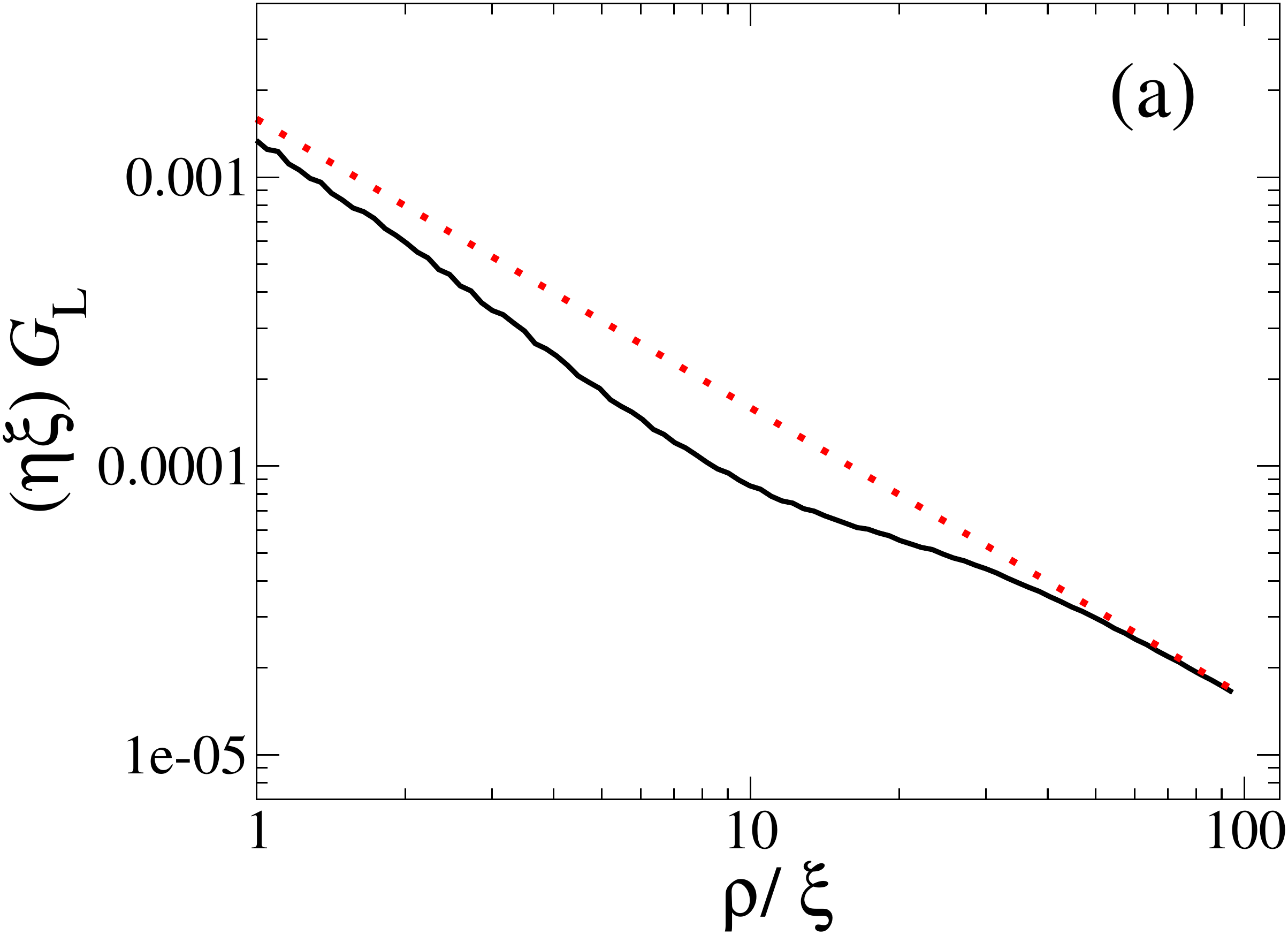}
\hspace{0.5cm} \includegraphics[width=0.45\linewidth]{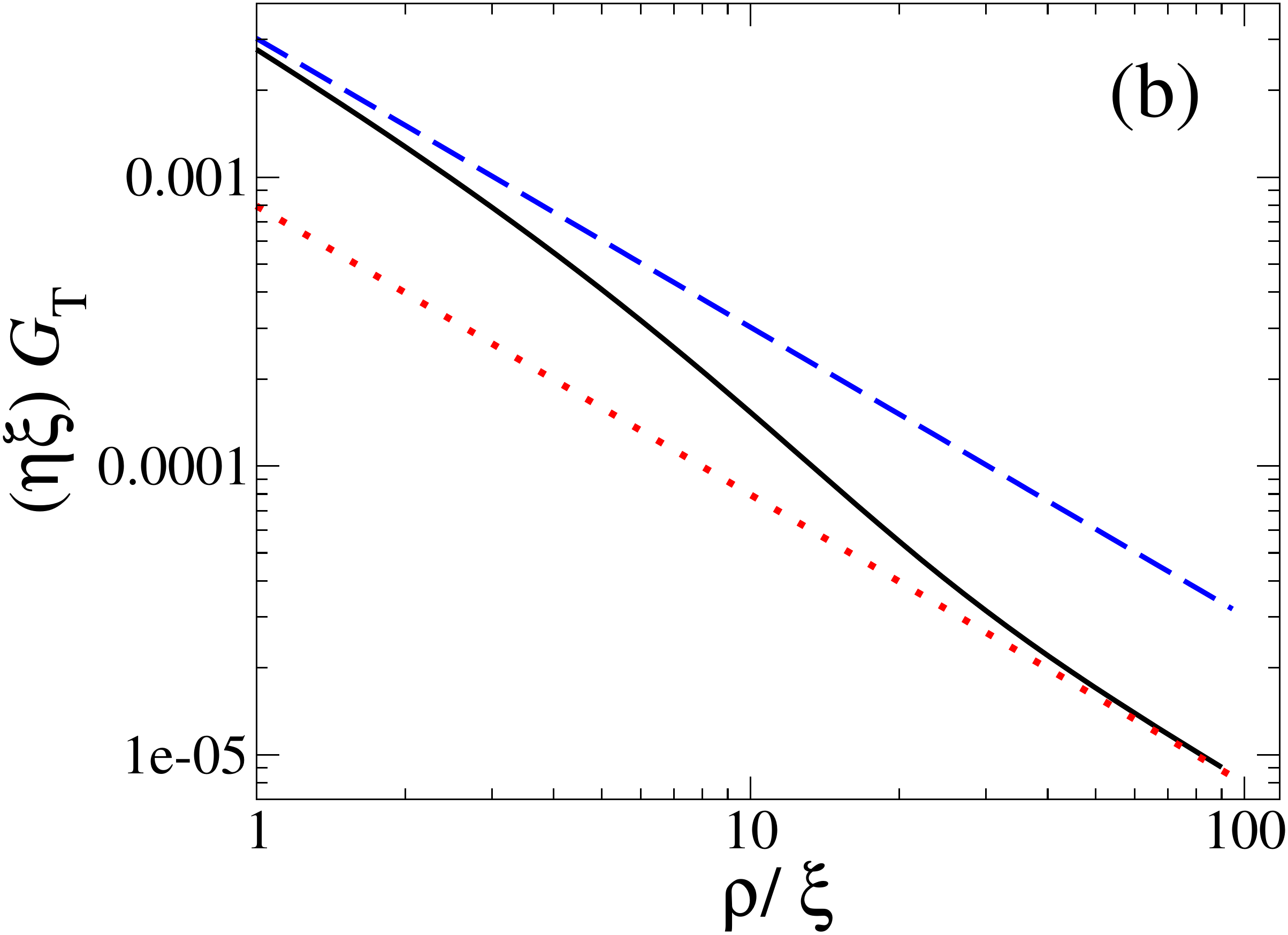}}
\caption{\setlength{\baselineskip}{6pt}
  Normalized longitudinal and transverse responses as a
  function of normalized separation. Once normalized, these functions
  depend on the parameters $\nu$ and $\etab/\eta$ alone. We have used
  $\nu=-1$ (in (a)), $\nu=0.1$ (in (b)), and $\etab/\eta=100$. The
  curves deviate from their asymptotic limits around
  $\rho\sim\lambda\gg\xi$. Panel (a) shows the longitudinal
  response. Numerical results (solid black curve) are shown along with
  the long- and short-distance asymptote (eq \ref{G_L_eq}; dotted red
  line). Panel (b) shows the transverse response. Numerical results
  (solid black curve) are shown along with the large-distance
  asymptote (dotted red line) and the short-distance one
  (eq~\ref{G_T_eq}; dashed blue line); see . The small Poisson ratios,
  corresponding to unrealistically compressible networks, are chosen
  to graphically emphasize the difference between the two regimes on a
  logarithmic scale. For a realistic value of $\nu=0.4$ we get a
  difference of about $2\%$ in (a) and $20\%$ in
  (b). Equations~\ref{G_L_eq} and \ref{G_T_eq} can be used to
  calculate the effect for any value of $\nu$. The value of
  $\etab$ is motivated by the actin networks of
  ref~\citenum{ViscoelasticIntermediate}, whose low-frequency response
  is governed by the viscous term, with viscosity about 100 times that
  of water.}
\label{fig_GLGT}
\end{figure}

From eqs~\ref{Gijr_eq_far} and \ref{Gijr_eq_near} we derive
two results which are particularly relevant for microrheology
experiments. The first is the longitudinal response,
\begin{equation}
\label{G_L_eq}
\rho\gg\lambda,\ \rho\ll\lambda: \ \ \ \ \
\GL(\rho) = {\cal G}_{xx}(\rho\hat{x})=\frac{1}{2\pi\etab \rho},
\end{equation}
where we have taken, without loss of generality, the direction
connecting the two points as the $x$ axis. Equation~\ref{G_L_eq}
corresponds to the case where both the perturbation and the response
are aligned with the direction between the two points. The second is
the transverse response,
\begin{equation}
  \GT(\rho)={\cal G}_{xx}(\rho\hat{y}) = {\cal G}_{yy}(\rho\hat{x})
  =\left\{\begin{array}{ll}
  \frac{1}{4\pi\etab\rho},\ \ \ \ \ \ &\rho\gg\lambda,\\
  \frac{1}{4\pi\etab} \frac{\beta^2\etab + \eta}{(\beta^2-1)\etab+2\eta}
  \,\frac{1}{\rho},&\rho\ll\lambda,
  \end{array}\right.
\label{G_T_eq}
\end{equation}
which corresponds to the case where the perturbation and the response
are perpendicular to the direction between the points. Having chosen
the $x$ axis along that direction, we have ${\cal G}_{xy}=0$ by
symmetry. Figure~\ref{fig_GLGT} presents the longitudinal and
transverse responses as calculated numerically, along with the two
asymptotic limits for small and large separations. Note the deviation
from the asymptotic behavior at distances $\rho\sim\lambda$, much
larger than $\xi$. We have selected a small value for the Poisson
ratio (a highly compressible network) to emphasize that deviation.

\begin{figure}[h]
\vspace{0.5cm}
\centerline{\includegraphics[width=0.45\linewidth]{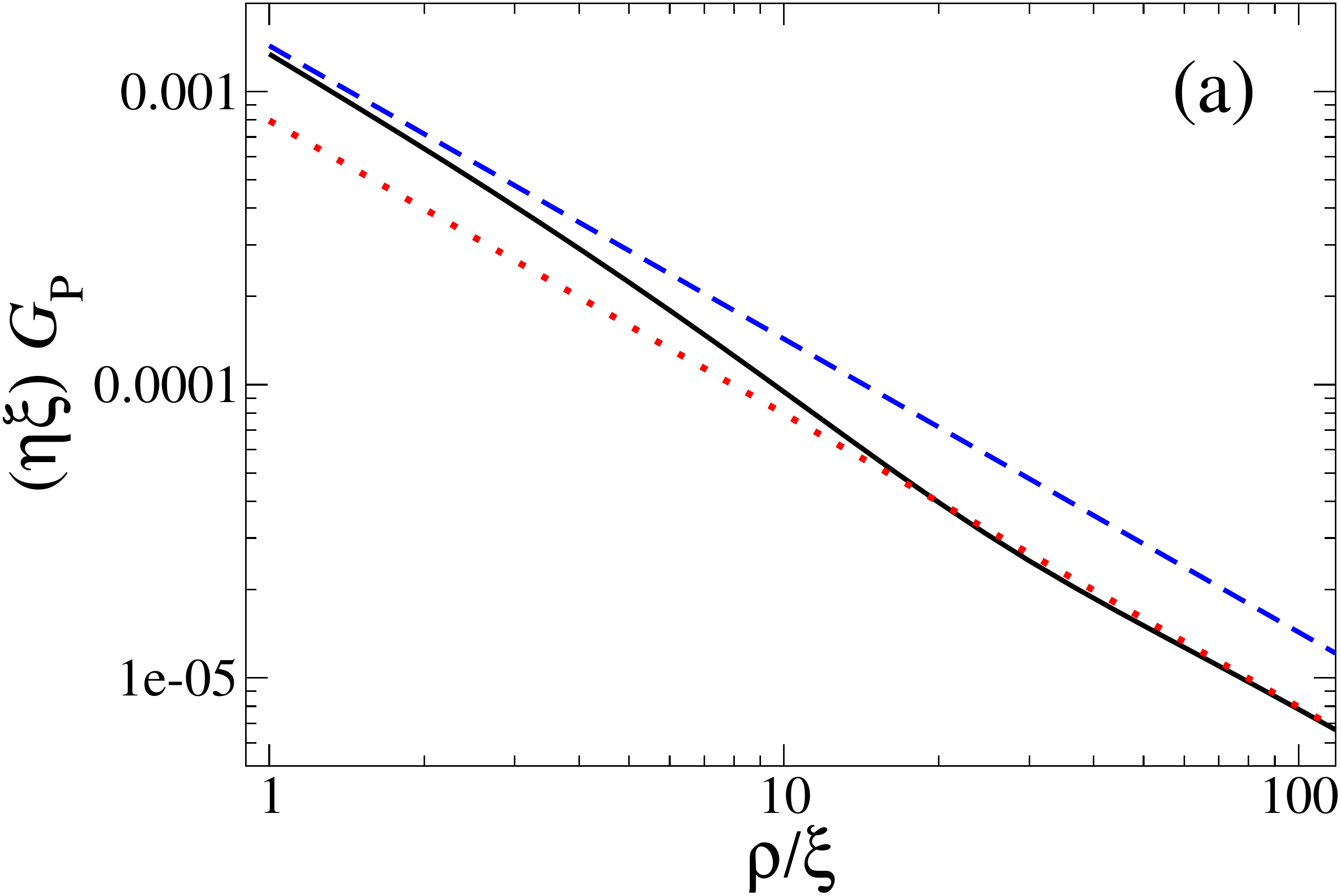}
\hspace{0.5cm} \includegraphics[width=0.45\linewidth]{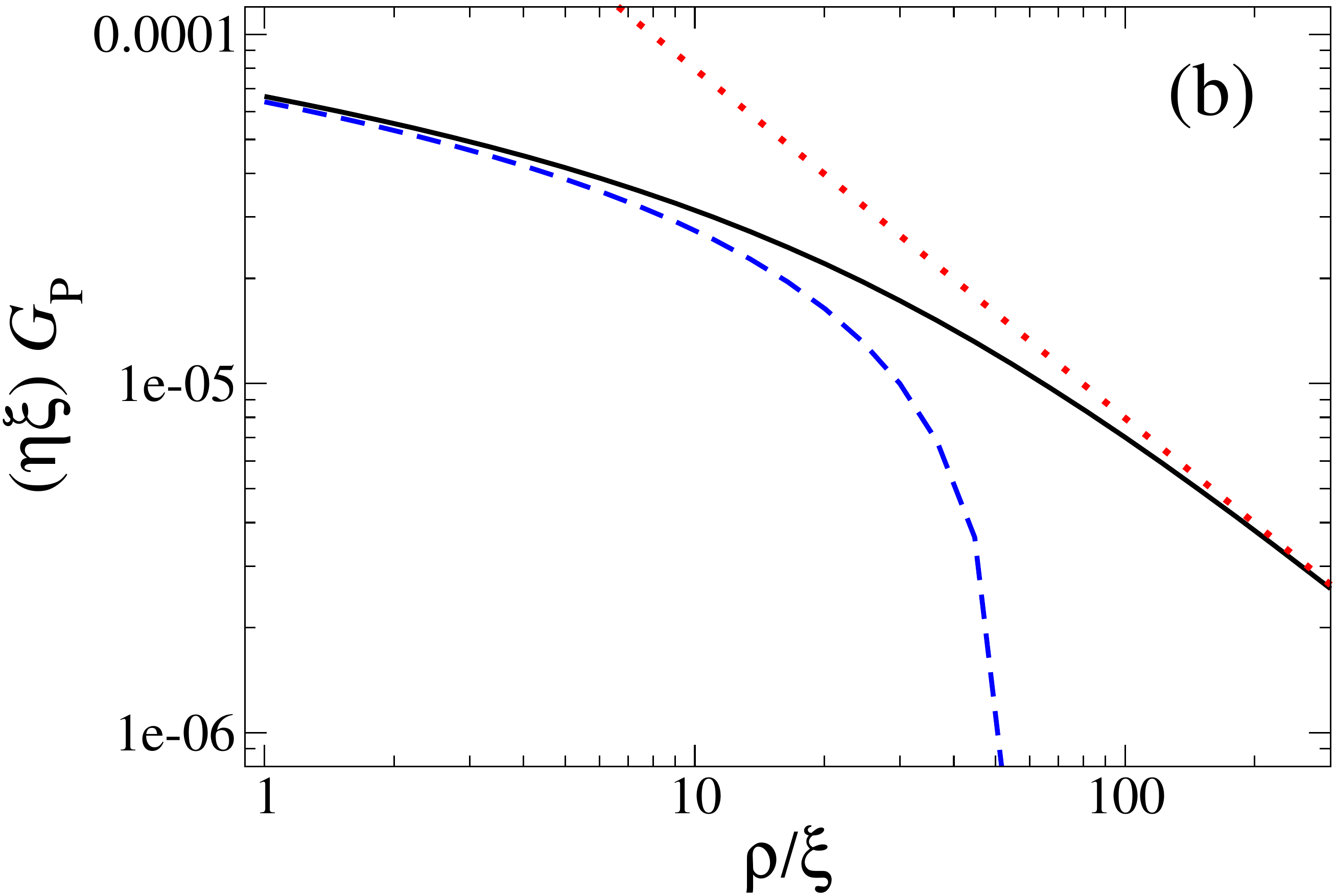}}
\caption{\setlength{\baselineskip}{6pt}
  Normalized perpendicular response as a function of normalized
  separation. Once normalized, these functions depend on the
  parameters $\nu$, $\etab/\eta$ and
  $\lec/\xi=\gamma/(i\omega\etab\xi)$. We have used $\nu=0.1$ and
  $\etab/\eta=100$. Panel (a) demonstrates the intermediate regime,
  taking $\lec=0$ (no surface tension). Numerical results (solid black
  curve) and the two asymptotes given by eqs~\ref{eq_GzzLargeDist}
  (dotted red line) and \ref{eq_Gzz_intermediate} (dashed blue line)
  are shown. Panel (b) focuses on the short-distance regime, where we
  have taken $\lec/\xi=100$. The numerical result (solid black curve)
  and the two asymptotes given by eqs \ref{eq_GzzLargeDist} (dotted
  red line) and \ref{eq_ln_2D} (dashed blue line) are shown. The small
  Poisson ratio, corresponding to an unrealistically compressible
  network, is chosen to graphically emphasize the difference between
  the two regimes on a logarithmic scale. For a realistic value of
  $\nu=0.4$ we get a difference of about
  $20\%$. Equations~\ref{eq_GzzLargeDist}--\ref{eq_Gzz_intermediate}
  can be used to calculate the effect for any value of $\nu$. The
  value of $\etab$ is motivated by the actin networks of
  ref~\citenum{ViscoelasticIntermediate}, whose low-frequency response
  is governed by the viscous term, with viscosity about 100 times that
  of water.}
\label{fig_Gzz}
\end{figure}

We now turn to the perpendicular response $\GP\left(\rho\right)={\cal
  G}_{zz}\left(\rho\right)$, where both perturbation and resulting
displacement are perpendicular to the surface.  The behavior of this
component is richer and affected by the surface tension $\gamma$. We
find three asymptotic behaviors. At distances larger than both the
elastic length $\lambda$ and the elasto-capillary length $\lec$ we
recover again the bulk response,
\begin{equation}
\label{eq_GzzLargeDist}
\rho\gg \max\left(\lambda,\lec\right): \ \ \ \ \ \GP(\rho)=\frac{1}{4\pi\etab \rho}.
\end{equation}
This result is independent of both $\lambda$ and $\lec$ and coincides
with the perpendicular response of the Boussinesq problem
\cite{LLelasticity} upon the appropriate replacements discussed
above. At distances much shorter than these length scales, we obtain
an effectively two-dimensional, logarithmic response, whose only
length scale is $\lec$,
\begin{equation}
\label{eq_ln_2D}
\rho\ll \min\left(\lambda,\lec\right): \ \ \ \ \
\GP(\rho)=\frac{1}{2\pi\etab\lec} 
\left[ \ln\left(\frac{\lec}{\rho}\right) - \gamma_E \right],
\end{equation}
where $\gamma_E\simeq 0.58$ is Euler's constant. This expression is
reminiscent of the flow response of fluid membranes, where the
elasto-capillary length plays here the role of the
Saffman-Delbr{\"u}ck length, acting as a two-dimensional cutoff
\cite{SaffmanDelbruk1975,Oppenheimer2009}.  At intermediate distances
we find another distinctive regime, which is unaffected by surface
tension and depends on the length scale $\lambda$ (through the Poisson
ratio $\nu$),
\begin{equation}
\lec\ll \rho\ll\lambda: \ \ \ \ \ \GP(\rho)=\frac{1}{4\pi\etab
  \rho}\frac{\beta^2}{\beta^2-1}.
\label{eq_Gzz_intermediate}
\end{equation}
where $\beta(\nu)$ has been defined in eq \ref{eq_def_beta}.  Figure
\ref{fig_Gzz} presents the spatial dependence of $\GP$, demonstrating
the three asymptotic regimes.

Examining the off-diagonal response ${\cal G}_{xz}(\rho)$, we find
that it vanishes for all $\rho$. The $xz$ component vanishes also in
the Boussinesq problem when the Poisson ratio is taken to be $1/2$
\cite{LLelasticity}. Thus the vanishing of this component is a result
of the overall incompressibility of the medium.

We would like to stress again two points. (a) In all the asymptotic
expressions for short distances, eqs~\ref{Gijr_eq_near}--\ref{G_T_eq} and
\ref{eq_ln_2D}, the distance $\rho$ must still be kept larger than
$\xi$ to ensure the validity of the continuum theory. This is why in
all figures we take $\rho/\xi>1$. (b) We give above the velocity
response; to obtain the displacement response one should divide the
results for ${\cal G}_{ij}$ by $i\omega$ and recall that $i\omega\etab=\Gb$.

\vspace{-0.5cm}
\section{Discussion}
\label{sec_discuss}

Let us summarize the main results which are relevant to
experiments. Equation \ref{eq_DR_result} gives the dispersion relation
of overdamped surface fluctuations as a function of wavevector in the
limits of small and large $q$, which can be probed by surface
scattering. To obtain the explicit decay rate one needs to know the
viscoelastic modulus of the specific material. We have demonstrated
this procedure for three simple frequency dependencies of the
modulus. Conversely, the small-$q$ limit can be used to extract
the bulk viscoelastic modulus if the surface tension is known, or vice
versa. The main novelty of the present work with respect to surface
scattering lies in the crossover between the two asymptotes, occurring
for $q\sim\lambda^{-1}$, as well as the large-$q$ asymptote, which
depends on the Poisson ratio $\nu$ (see Figure~\ref{fig_DR}). These
features, which depend on the intrinsic length of the material, have
not been considered by earlier theories. For example, they may be used
to extract the compression modulus of the bare polymer network. This
modulus has eluded measurement, because decoupling it from the total
compression modulus of the medium (which is governed by the solvent's
very large modulus) is hard \cite{HaimEPJE}. The dispersion relation
of eq \ref{eq_DR_result} is relevant to low frequencies where inertia
is negligible. Comparison with refs~\citenum{HardenPincus1st,Pincus},
which included inertial effects, reveals that this requirement is
fulfilled for $\omega\ll\min(\sqrt{\gamma q^3/\rho_{\rm m}},\etab
q^2/\rho_{\rm m})$, where $\rho_{\rm m}\sim 1$ gr/cm$^3$ is the
medium's mass density. Checking against eq \ref{eq_DR_result}, we find
that the dispersion relation is consistent for $q\gg\gamma\rho_{\rm
  m}/\etab^2$. Thus, the range of validity increases with decreasing
surface tension and increasing viscoelasticity. For $\gamma\sim 10$
erg/cm$^2$ and $\etab\sim 1$ poise, the requirement is $q\gg 10$
cm$^{-1}$ which generally holds. However, at sufficiently large $q$
(i.e., large $\omega$), the bulk viscosity will no longer be much
larger than the solvent's, and the requirement above for $\omega$ will
not be valid.

Our predictions concerning the response to a localized force are
relevant to two-point microrheology \cite{TwoPointCrocker}. Equations
\ref{G_L_eq} and \ref{G_T_eq} give the longitudinal and transverse
velocity responses at the surface. Multiplied by $k_{\rm
  B}T/(i\omega)$, they give the correlations between the displacement
fluctuations of two tracer particles lying on the surface and
separated by a distance $\rho$, along and transverse to the separation
vector, respectively. The results for large separations are found to
be equivalent to those for an elastic medium \cite{LLelasticity}. This
was experimentally observed for thick films made of hyaluronic acid
gels \cite{LadamSackmann}. However, once the separation becomes of
order $\lambda$ or less, we expect appreciable deviations from the
elastic result (see Figure \ref{fig_GLGT}). We note that $\lambda$ is
typically much larger than the mesh size $\xi$. This prediction,
assuming that the experiment covers sufficiently large distances
between tracer particles, can be used to measure $\lambda$ and extract
the network's compression modulus.

In the direction perpendicular to the surface we have found a richer
behavior arising from the dependence on the elasto-capillary
length. This includes a quasi-two-dimensional membrane-like behavior
at small distances (eq \ref{eq_ln_2D}), and a distinctive intermediate
regime which depends on the network's compressibility (eq
\ref{eq_Gzz_intermediate}). Unfortunately, it should be hard to
measure displacements perpendicular to the surface at sufficient
spatial and temporal resolutions.

The predictions concerning correlations between surface displacements
can be used to perform ``non-invasive'' microrheology, where
the viscoelastic moduli of the medium are measured by tracking
particles on the surface
\cite{LadamSackmann,expDennin2012,expDennin2013,expDennin2014,KomuraMaterials2012,KomuraEPL2012}. Importantly,
at sufficiently large distances, this kind of two-point microrheology
should be insensitive to surface heterogeneities. It would be
insensitive also to features of the particle-surface interactions such
as the possible meniscus around the probe particles.

Underlying our results is the assumption of finite friction between
network and solvent, which departs from the strong-coupling limit
considered earlier \cite{HardenPincus1st,Pincus,Pleiner1988}. Such
friction was found necessary to account for experiments on entangled
actin networks
\cite{ViscoelasticIntermediate,DynamicCorrelationLength}. It is
expected to occur, in addition, in semi-dilute solutions of long
polymers and close to a critical (\eg theta) point, where the network
is highly compressible.

In this work we have highlighted the spatial properties of the surface
response. The results can be alternatively discussed from the
perspective of frequency dependence. Such a discussion requires
detailed knowledge of the viscoelastic properties of the material (\ie
$\Gb(\omega)$), which goes beyond the generic theory presented
here. In the examples provided above we have assumed the dominance of
the viscous contribution at low frequency, as is known for actin
networks \cite{ViscoelasticIntermediate}. In other limits the response
would be quite different and its treatment would require specific
details. For example, the elasto-capillary length $\lec\sim
1/\Gb$, whereas the compressibility length $\lambda\sim
\sqrt{\Gb/\omega}$. This implies a delicate interplay between the
different effects as a function of frequency.

The treatment of a two-fluid boundary could be made more precise in
the future. Here we have assumed that the two components move together
at the surface (boundary condition eq \ref{bc_freeSurface1}). Such
strong coupling should be valid when both components are strongly
repelled from the outer phase, resulting in a sharp interface of large
surface tension. To relax this assumption one should replace boundary
conditions \ref{bc_freeSurface1} and \ref{bc_freeSurface2} by two
stress-balance conditions for the two components separately. These
conditions would include a friction term proportional to the
difference in the surface velocities of the two components. Another
assumption behind boundary condition \ref{bc_freeSurface2} is that the
external force is applied to both components. This will be the case in
the common microrheology scenario where the force is applied to an
inert bead much larger than $\xi$ and in physical contact with both
network and solvent. There may be other cases where, for example, the
force is applied by a particle bound to the network alone. The same
refinement described above would allow also to treat such
scenarios. The two stress-balance conditions can include separate
external forces for the two components.

The present work has addressed the surface response of an indefinitely
thick sample. Considering a film of finite thickness will introduce
another length scale, which should lead to even richer behavior. In
particular, since many of the results presented here depend on the
length $\lambda$, which may be orders of magnitude larger than the
network's mesh size in the case of low network compressibility, a
strong effect of the finite thickness is expected. This will be
addressed in a forthcoming publication.

\vspace{-0.5cm}
\begin{acknowledgement}
This work has been supported by the Israel Science Foundation (Grants
No.\ 164/14 and No.\ 986/18).
\end{acknowledgement}

\vspace{-0.5cm}
\bibliography{paper2_bibl_cm}

\end{document}